\PassOptionsToPackage{hidelinks}{hyperref}

\documentclass[a4paper,twocolumn,11pt,accepted=2026-06-17]{quantumarticle}
\pdfoutput=1

\usepackage{hyperref}

\usepackage[T1]{fontenc}
\usepackage[utf8]{inputenc}
\usepackage[english]{babel}

\usepackage{amsmath,amssymb}
\usepackage{braket} 

\usepackage{booktabs,tabularx,makecell}
\newcolumntype{Y}{>{\centering\arraybackslash}X}

\usepackage{tikz}

\usepackage[
  backend=biber,
  style=numeric,
  sorting=none,
  doi=true,
  url=true,
  eprint=true
]{biblatex}
\begin{document}

\title{Controller-decoder system requirements derived by implementing Shor’s algorithm with surface codes}

\author{Yaniv Kurman}
\affiliation{Quantum Machines Inc., Tel Aviv, Israel}
\orcid{0000-0002-5784-3573}
\author{Lior Ella}
\affiliation{Quantum Machines Inc., Tel Aviv, Israel}
\author{Nir Halay}
\affiliation{Quantum Machines Inc., Tel Aviv, Israel}
\author{Oded Wertheim}
\affiliation{Quantum Machines Inc., Tel Aviv, Israel}
\author{Yonatan Cohen}
\affiliation{Quantum Machines Inc., Tel Aviv, Israel}

\maketitle

\begin{abstract}
  Quantum Error Correction (QEC) is regarded as the most promising path to quantum advantage. The success of QEC relies on achieving quantum gate fidelities below the error threshold of the QEC code, while accurately decoding errors through classical processing of the QEC stabilizer measurements. In this paper, we uncover the critical system-level requirements from a controller-decoder system (CDS) necessary to successfully execute the next milestone in QEC: a non-Clifford circuit. Using a representative non-Clifford circuit, of Shor’s factorization algorithm for the number 21, we convert the logical-level circuit to a QEC surface code circuit and finally to the physical level circuit. By taking into account realistic implementation aspects using typical superconducting qubit processor parameters, we reveal a broad range of core requirements from any CDS aimed at performing error corrected quantum computation. Our findings indicate that the controller-decoder closed-loop latency must remain within tens of microseconds, achievable by distributing decoding data into several decoders while ensuring fast communication between decoders and with the controller. By extending existing simulation techniques, we simulate the complete fault-tolerant factorization circuit at the physical level, demonstrating that near-term hardware performance in the scale of 0.1\% physical error rates and 1000 qubits, are sufficient for a successful circuit execution. Overall, the requirements outlined here set the stage for near- and medium-term experimental realizations of non-Clifford QEC circuits.
\end{abstract}

\section{Introduction}
Quantum error correction (QEC) offers a concrete route for reaching practical quantum computation through its ability to suppress logical quantum errors by orders of magnitude, provided the physical quantum error rates are below a certain threshold \cite{dennis2002topological, knill1998resilient}. As quantum hardware scales from a few qubits to hundreds, and even thousands, experimental demonstrations have begun to reach critical QEC milestones. These milestones range from extending the lifetimes of bosonic codes within a single optical cavity  \cite{sivak2023real, lachance2024autonomous} to multi-qubit stabilizer-code demonstrations \cite{kelly2015state, krinner2022realizing}. The stabilizer codes have shown great success in suppressing logical idle errors when increasing number of qubits \cite{google2023suppressing,google2025quantum}, and running fault-tolerant shallow Clifford circuits \cite{bluvstein2024logical,ryan2404high, hetenyi2024creating, paetznick2024demonstration}.

Two classical hardware components determine the performance of an error-corrected quantum computer:
(i) the controller, which executes the quantum logic, is vital in minimizing the physical gate error \(P_{\mathrm{phys}}\) for a given quantum hardware.
First, the controller’s analog pulse-generation capabilities, noise levels, and stability significantly affect gate fidelity and qubit coherence times~\cite{vepsalainen2022improving, klimov2020snake}.
Second, the controller's ability to run calibrations efficiently, quickly, and frequently allows continuous optimization and stabilization of the fidelities to achieve higher average fidelities over time~\cite{kelly2016scalable, xu2022automatic, proctor2020detecting}.
(ii) the quantum error decoder, which classically processes the physical measurements of the QEC sequence to detect quantum errors\cite{terhal2015quantum}.
After decoding, the QEC logical error rate follows the general scaling formula~\cite{fowler2012surface},
\begin{equation}
  P_{\mathrm{log}} \propto \left(\frac{P_{\mathrm{phys}}}{P_{\mathrm{th}}}\right)^{(d+1)/2},
  \label{eq:log-scaling}
\end{equation}
where \(d\) is the QEC code distance and \(P_{\mathrm{th}}\) is the QEC code error threshold.

The error threshold, \(P_{\mathrm{th}}\), is governed primarily by the chosen QEC code, but also by the decoding algorithm and the prior information it uses. Surface-code implementations have demonstrated remarkably high thresholds of order \(P_{\mathrm{th}}\sim\!1\%\) ~\cite{dennis2002topological, fowler2012surface}, with variances according to the decoder's knowledge about the quantum hardware ~\cite{google2023suppressing, google2025quantum} and the specific decoding strategy employed (showing tradeoff between the accuracy and complexity) ~\cite{demarti2024decoding}. Overall, reducing logical errors is possible by increasing the code distance (adding qubits) only if \(P_{\mathrm{phys}} < P_{\mathrm{th}}\), that is, only if across the up-scaled quantum processor the controller maintains a low average \(P_{\mathrm{phys}}\) and the decoder preserves a high \(P_{\mathrm{th}}\).

Reaching these controller and decoder requirements does not ensure a system's ability to support quantum computation with QEC. Executing non-Clifford circuits with QEC, essential for achieving quantum advantage~\cite{gottesman1998heisenberg}, necessitates incorporating a decoding-dependent gate (feed-forward) into the circuit for each non-Clifford gate~\cite{fowler2012surface, boykin1999universal}. Until the decoding result is reached and these gates are implemented, additional to-be-decoded data is accumulated during QEC idling. Therefore, QEC circuits composed from a set of non-Clifford gates create new requirements on the real-time performance of the controller, decoder, and their integration. 

To date, the primary decoder requirement has been to maintain a decoding rate faster than the QEC data-generation rate~\cite{terhal2015quantum}. This throughput requirement is the benchmark for state-of-the-art decoders~\cite{google2025quantum, liyanage2023scalable, barber2025real, battistel2023real} and a driving factor for QEC-focused controller architectures~\cite{fu2019control}, decoder architectures~\cite{das2022afs, maurya2024managing}, and system architectures \cite{wu2024lego}. In a recent design~\cite{maurya2024managing}, the authors showed how the memory requirements, the decoding latency, and the logical error rates may vary as the data to decode increases, going beyond previous high-level resource estimation approaches \cite{beverland2022assessing}. Finally, preliminary real-time benchmarks for a combined controller–decoder system have recently been proposed~\cite{kurman2025benchmarking}, with real-time, low-latency feedback recently demonstrated experimentally on a superconducting processor~\cite{caune2024demonstrating}.

However, a comprehensive analysis that is based on an end-to-end breakdown of a concrete circuit has been lacking, making it difficult to specify requirements and identify critical bottlenecks for the next experimental QEC milestone. In particular, gaps remain in understanding precise latency requirements—the time between the last measurement that can flip a decoding-dependent feed-forward gate and the gate's execution. In addition, it is unclear what are the effects of exceeding these latency limits, the required real-time control-flow operations, and the sizes of the decoding graphs, number, and internal connectivity requirements in order to support non-Clifford QEC circuits.

In this paper, we present an end-to-end analysis of the controller–decoder system (CDS) requirements based on the surface-code implementation of a 5-qubit factorization circuit for the number 21 (Fig.~\ref{fig:illustration}; circuit derived in Appendix~\ref{app:circuit}). Based on this representative non-Clifford circuit, we perform a detailed step-by-step breakdown to establish concrete CDS requirement. First, we map the logical circuit to a surface-level circuit comprised by 18 surfaces when assuming nearest-neighbor connectivity constraints. The 5 computational surfaces are dynamically teleported while ancillary surfaces are used for fault-tolerant logic. We then compile the surface-level circuit to a physical-level circuit in an end-to-end physical-level simulation, showing that physical errors of $0.1\%$ and a surface distance of $d=5$ (1000 physical qubits) are sufficient for reaching meaningful logical results. The physical-level simulations allowed extracting the data flow requirements as well as the controller's front-end execution requirements.

\begin{figure*} [!t]
\includegraphics[width=\textwidth]{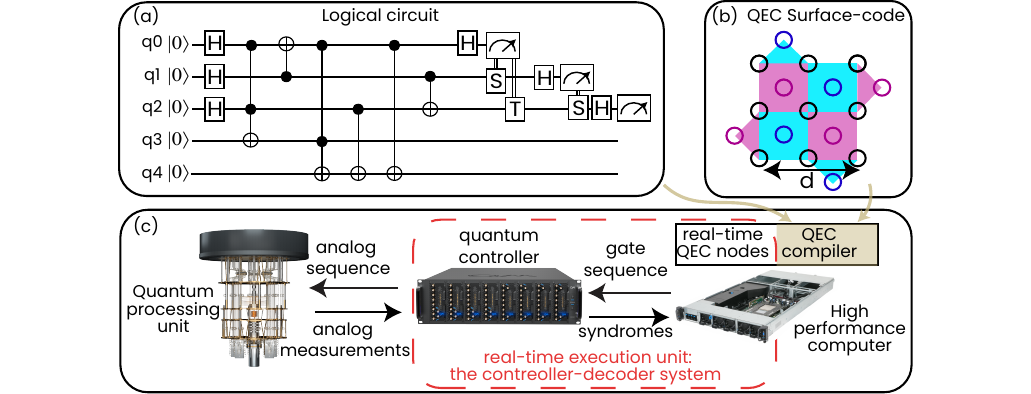}
 \caption{\label{fig:illustration} Controller-decoder system (CDS) for executing non-Clifford QEC circuits.
(a) The examined logical circuit which factorizes the number \(21\).
(b) A distance-3 surface-code layout. Data qubits (black) store the logical information; ancilla qubits (blue, pink) perform stabilizer measurements.
(c) Control and decoding stack: a high-performance computer (HPC) compiles the logical circuit and chosen QEC code into real-time instructions for the controller-decoder system (CDS, dashed red), comprising the quantum controller and low-latency HPC nodes implementing the decoder. During execution, the controller drives the QPU, streams stabilizer outcomes (syndromes) to the decoder, and applies feed-forward updates to subsequent gates as a result of real-time decoding. The resulting real-time data flow and processing set explicit requirements on CDS latency, throughput, resources, and structure, needed for surface-code-protected non-Clifford circuits.
}
\end{figure*}

To analyze the decoding aspects, we group syndrome measurements into separate decoding tasks according to the feed-forward operations they affect. Having more than a single non-Clifford gate create multi-task dependencies where the result and feed-forward time of one task modify the size of subsequent tasks \cite{kurman2025benchmarking}. In our concrete example, the surface-level circuit includes 13 decoding tasks, with a parallel processing requirements of up to 4 decoding tasks. We analyzed the number of syndromes in each decoding task and their feed-forward latency as a function of the physical error rate and the code distance. In addition, we find that the additional error due to delayed feed-forward is not dominant if it remains within tens of microseconds throughout the circuit (assuming superconducting quantum hardware), which is expected to be the case until utility-scale fault-tolerant computation. 

Although physical error rates of $0.1\%$ enable a more accurate execution of our circuit without QEC, the conceptual requirements and underlying logic described here remain relevant once QEC surpasses near-intermediate-scale-quantum (NISQ) computation. While our derivation focuses on a specific example, the analysis sets clear requirements on each component of the CDS, details the inter-decoder and controller-decoder integration requirements, as well as the control-flow commands. These practical set of specifications for controller-decoder systems in  QEC circuits is a foundational framework for advancing towards scalable fault-tolerant quantum computation.

\section{Non-Clifford surface code circuits}
\label{sec:surface_gates}
\begin{figure*} [!t]
\includegraphics[width=\textwidth]{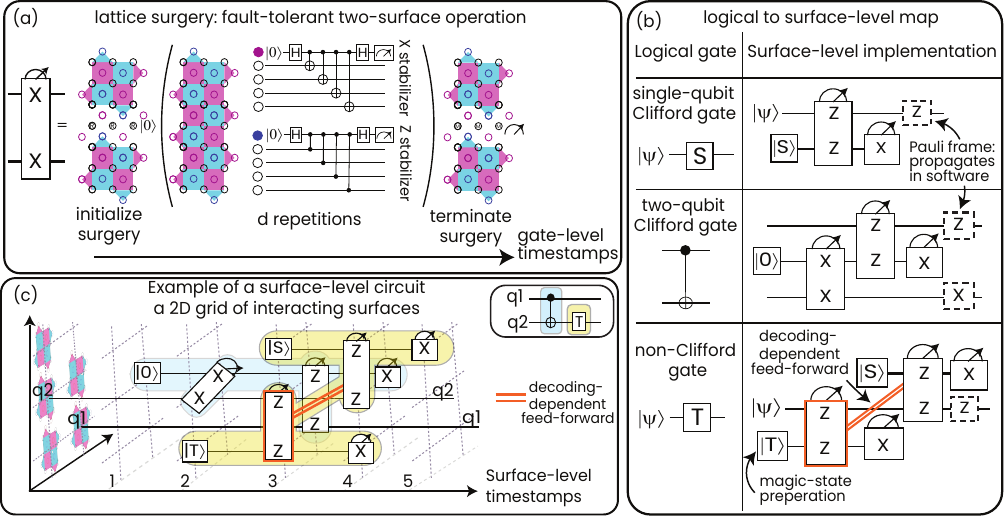}
\caption{\label{fig:logical_to_surface} Surface-code quantum computation. (a) Fault-tolerant (FT) two-surface gates are performed via lattice surgery, i.e., measurements the parity in the \(Z/X\) bases between multiple surfaces. Shown is an \(XX\) parity measurement using a line of ancillary qubits which are initialized, take part in \(d\) stabilizer rounds (depth-8 circuit) on the merged surface, and then measured to terminate the surgery. (b) Logical surface-level building blocks: FT initializations/measurements along logical Pauli axes and lattice surgeries, and a non-Clifford gate implemented by a magic-state preparation with decoder-dependent feed-forward. We chose a \(T=\mathrm{diag}(1,e^{i\pi/4})\) gate, using a \(\ket{T}=\ket{0}+e^{i\pi/4}\ket{1}\) state that can be prepared A non-FT within \(d\) rounds or with higher fidelity using distillation or cultivation at larger space--time cost. In this $T$-gate implementation, the feed-forward (e.g., an \(S\)-gate correction) must be applied before the next non-commuting gate. In all types of surface codes gates, measuring an ancillary surface modifies a computational qubit Pauli frame (dashed). (c) Surface-level circuit example: a logical CNOT (cyan) followed by a \(T\) gate (yellow) using five surfaces. Each timestamp equals \(d\) stabilizer rounds. In this example, all surfaces are active at timestamps 3 and 4. The \(q_2\) and \(\ket{S}\) surfaces continue to idle (via stabilizer rounds) before their decoding-dependent surgery, until the controller knows the decoding outcome of the yellow \(ZZ\) at 3.}
\end{figure*}

The first step in deriving the control procedure for the factorization circuit is to convert the original logical circuit into a logical circuit which uses the native gate set of the QEC code, thereby maximizing fault-tolerant gates. In this section, we explain the gate set for the surface code, how these gates are implemented, and what are their high-level requirements on the CDS.

\subsection{Clifford gates with surface codes}
\label{sec:Clifford}
Fault-tolerant (FT) logical gates are defined in the context of QEC as gates that obey the relation of Eq.~\eqref{eq:log-scaling}: their logical error rate is exponentially reduced as the code distance increases. In the surface code, the basic FT block corresponds to $d$ stabilizer rounds of a distance-$d$ surface. The single-surface FT gates include (i) initialization in the $X$ or $Z$ basis, implemented by data qubit initialization followed by $d$ stabilizer rounds, (ii) measurement in the $X$ or $Z$ basis, implemented by measuring the data qubits after $d$ stabilizer rounds, and (iii) idling, implemented by $d$ stabilizer rounds. After decoding the stabilizer measurements, these three gates have a similar logical error of a single FT block. Recent experiments have showed the FT of these protocols~\cite{google2023suppressing, google2025quantum}.

The fundamental FT two-surface gate, which is the central building block of surface-code logic~\cite{chamberland2022universal, litinski2019game, higgott2023improved} is the lattice surgery~\cite{horsman2012surface}, shown in Fig.~\ref{fig:logical_to_surface}a. This process implements the parity measurement between two or more surfaces in the $Z/X$ basis (e.g., $ZZ$, $XX$, or $ZX$ for two surfaces). A logical measurement outcome of $0$ corresponds to even parity (e.g., collapsing into the $\ket{00}$ and $\ket{11}$ subspace for a $ZZ$ surgery). In the example in Fig.~\ref{fig:logical_to_surface}a, an $XX$ measurement between two distance-3 surfaces is done by converting them into an elongated surface for $d$ rounds, where the logical $X$ edges of the two surfaces are patched together. This is implemented by initializing a line of ancillary qubits located between the relevant edges and adding four $X$-stabilizers whose joint parity corresponds to the logical $XX$ operator. This operation is FT because the logical $XX$ is probed for $d$ rounds, and it is terminated by measuring all ancillary qubits. After decoding, the two-surface lattice surgery has a logical error rate equivalent to that of two FT blocks plus an additional contribution from the relative patching volume.

These FT single-surface and multi-surface gates in the $Z/X$ basis are sufficient to implement Clifford computation along the $Z$ and $X$ axes. For example, a CNOT between two surfaces is implemented using an additional ancillary surface and two lattice-surgery operations (Fig.~\ref{fig:logical_to_surface}b, middle)~\cite{horsman2012surface}. Notably, the three logical measurements in this construction, and in fact all projective measurements appearing in surface-code computation, are intrinsically random and yield '0' or '1' with equal probability, where a '1' measurement outcome implies that the computational qubits experienced a byproduct gate. Within a Clifford circuit, the measurement outcomes from a lattice surgery or from an ancillary surface also determine whether a logical Pauli byproduct has been applied to the participating computational qubits (dashed squares in Fig.~\ref{fig:logical_to_surface}b). Importantly, these Pauli byproducts do not require physical correction. Instead, they are tracked in software as Pauli-frame updates, which may only flip the outcomes of future logical measurements (for example, a logical $X$ frame will flip an $Z$-basis measurement outcome)~\cite{fowler2012surface}. To complete the Clifford gate set, the $X^{1/2}$ and $S$ gates (Fig.~\ref{fig:logical_to_surface}b, top) rely on logical initialization of an $\ket{S} = \ket{0} + i\ket{1}$ state, which was recently shown to admit a fault-tolerant preparation procedure, albeit with a higher error rate than the $Z/X$ initializations~\cite{gidney2024inplace}.

In terms of CDS requirement, implementing Clifford circuits with surface codes (using only the gate set described above) mainly relies on maintaining $P_{\mathrm{phys}} < P_{\mathrm{th}}$. The gate sequence of the entire circuit is deterministic, all byproduct operations are Pauli-frame updates that can be propagated to the end of the computation, and decoding can in principle be deferred until after the circuit executing has finished. In other words, a slow decoder or controller–decoder communication channel does not accumulate additional logical errors, it only delays the circuit outcome. This picture fundamentally changes once non-Clifford gates are introduced.

\subsection{Non-Clifford gates with surface codes}
\label{sec:non-Clifford}

Non-Clifford gates are required to complete a universal gate set and introduce an additional, critical set of requirements for the CDS. The standard route is to implement them via a preparation of a \emph{magic state}~\cite{bravyi2005universal} (a state outside the cardinal axes of the Bloch sphere), such as $\ket{T} = \ket{0} + e^{i\pi/4}\ket{1}$. The key advantage is that, once the magic state has been prepared, all subsequent operations needed to enact the corresponding non-Clifford gate can be FT~\cite{bravyi2005universal, fowler2012surface}. Specifically, the byproduct operation resulting from a logical measurement outcome of $1$ in a lattice-surgery implementation of the $T$ gate is a Clifford gate; however, this Clifford correction cannot be absorbed into the Pauli frame. If a logical $1$ is obtained, which is only known after decoding, the corresponding Clifford correction must be physically applied (as illustrated in the $T$-gate implementation in the bottom panel of Fig.~\ref{fig:logical_to_surface}b). 

This requirement is the origin of a crucial CDS constraint in non-Clifford circuits: the need for decoding-dependent feed-forward. The implementation shown in Fig.~\ref{fig:logical_to_surface}b (right) allows this feed-forward to be delayed until after any commuting gates in the surface-level circuit, thereby relaxing the requirement on the feed-forward latency (the time between the last measurement in the decoding task and the feed-forward pulse that depends on it). If the feed-forward is not applied in time, syndrome measurements that still need to be decoded continue to accumulate, since $\ket{\psi}$ must remain idle. As a consequence, when running non-Clifford circuits the CDS must support a fast controller–decoder round trip as well as fast, high-throughput decoding~\cite{terhal2015quantum, kurman2025benchmarking}. Notably, in alternative $T$-gate constructions, such as the ``$\pi/4$-autocorrected'' scheme~\cite{litinski2019game}, the feed-forward instead determines the measurement basis of an ancillary surface, and the required correction can be implemented at any point prior to the decoding-dependent feed-forward of the subsequent non-commuting gate.

An example of a small segment of a non-Clifford circuit is shown in Fig.~\ref{fig:logical_to_surface}(c), which implements a logical CNOT (cyan) followed by a $T$ gate (yellow) using  the primitives from Fig.~\ref{fig:logical_to_surface}(b). The circuit includes five surface-level timestamps (each $d$ stabilizer rounds) and  five surfaces arranged on a two-dimensional grid: two computational surfaces, denoted $q1$ and $q2$, and three ancillary surfaces. The example shows how logical gates are parallelized. At timestamp 2 an ancillary surface performs an $XX$ lattice surgery with the computational surface $q2$ while a $\ket{T}$ state is initialized on another ancillary surface as part of a T-gate on $q1$. This initialization can be performed within $d$ stabilizer rounds without additional time or qubit overhead via state injection, or the state can be teleported from extra ancillary surfaces after preparation using any of the methods discussed below. 

Notably, the decoding outcome of the yellow $ZZ$ measurement at timestamp 3 (associated with the non-Clifford gate) determines whether the lattice surgery at timestamp 4 is executed. This decoding processes all previous syndrome measurements that can flip the effective $ZZ$ outcome at timestamp 3, including physical errors during the $XX$ surgery at timestamp 2 and during the $\ket{0}$ initialization. In the schematic, the feed-forward is depicted as being applied immediately after the yellow $ZZ$ measurement in timestamp 3. In practice, decoding is not instantaneous, so the $q2$ surface and the surface initially prepared in the $\ket{S}$ state must idle under stabilizer rounds until the CDS executes the decoding decision. These additional stabilizer rounds generate extra syndromes that must also be decoded to determine the feed-forward for the next non-Clifford gate or the final circuit outcome. This idling-induced desynchronization between logical qubits' syndrome-extraction cycles, and policies to mitigate it, have recently been studied directly~\cite{maurya2025synchronization}.

Apart from the requirement of fast decoding-dependent feed-forward, a second major challenge in running non-Clifford circuits is the magic-state initialization. These initializations in surface codes are non-FT in the sense that increasing the code distance does not exponentially suppress their logical error, even when $P_{\mathrm{phys}} < P_{\mathrm{th}}$, because no two-dimensional stabilizer code admits a transversal universal gate set~\cite{eastin2009restrictions}.

Several techniques address the magic-state initialization challenge, trading target initialization error against time or qubit overhead. Magic-state injection~\cite{li2015magic, gidney2023cleaner} is the most resource-efficient approach, but it yields the highest error rate. Injection is performed in \(d\) stabilizer rounds and produces a logical \(\ket{T}\)-state with an infidelity that scales linearly with the physical error rate \(p\). Zero-level distillation is a method that applies a small error-detecting code before initialization, improving the scaling to \(O(p^{2})\) at modest additional cost~\cite{itogawa2025zerolevel}. Magic-state distillation reaches the lowest initialization errors, with infidelity scaling as \(O(p^{r})\), where increasing \(r\) arbitrarily is done by concatenating distillation protocols~\cite{litinski2019magic, gidney2019efficient}. However, distillation requires substantial space--time resources, making it impractical in the near-term and even medium-term. A recent approach called magic-state cultivation, achieves intermediate target infidelities between injection and distillation by starting from injection and gradually increasing the logical code distance through multiple verification stages~\cite{gidney2024magic}. The resulting fidelity can be tuned down to \(10^{-10}\) (for \(p=10^{-3}\)), depending on the chosen discarding thresholds. Choosing between these methods is primarily dictated by the target initialization error and the available physical-qubit and time budgets. Resource estimates for large-scale fault-tolerant quantum computation show that magic-state preparation is often the dominant contributor to total algorithm cost~\cite{fowler2012surface, litinski2019game}, and that reducing this overhead can lower the required qubit count by more than an order of magnitude~\cite{gidney2025factor}.

From the perspective of CDS requirements, magic-state injection, zero-level distillation, and cultivation all rely on fast, real-time measurement-based branching at the controller level. The controller must support low-latency conditional operations (``if--else''-style flow control) to implement repeat-until-success protocols efficiently, rather than relying on purely offline post-selection after the full circuit execution, which is prohibitively inefficient at scale. Cultivation additionally involves real-time measurement-dependent feed-forward and decoding-dependent branching. Measurement-dependent feed-forward requires sub-100-ns latencies to correct local stabilizers, whereas the decoding-dependent discarding decisions can be deferred until just before the state is consumed. Finally, magic-state distillation can be viewed as a full non-Clifford circuit that includes injected or cultivated \(\ket{T}\) states, FT Clifford gates, and decoding-dependent feed-forward.

\section{The surface-level circuit}
\label{sec:fact_circuit}
In this section, we present the steps needed to convert the target logical quantum circuit into a full circuit constructed by surface-code operations, i.e., a surface-level circuit. That is, to implement the target logical circuit using the building blocks described in Sec.~\ref{sec:surface_gates}.

\subsection{Surface-compatible logical circuit}
\begin{figure*} [!t]
\includegraphics[width=\textwidth]{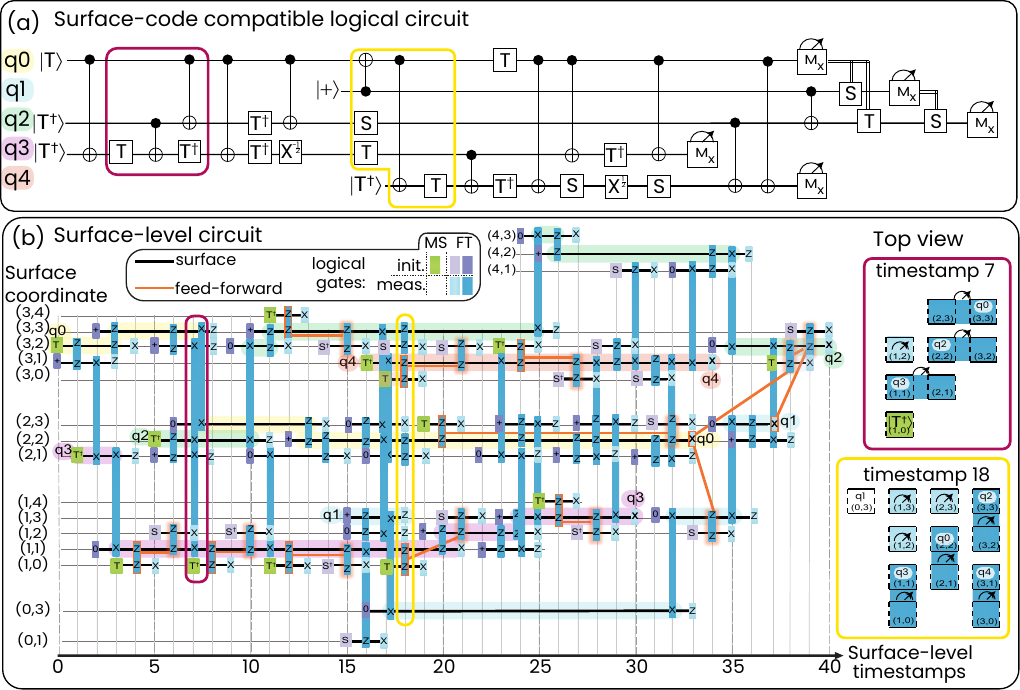}
 {\caption{\label{fig:surface_level} The surface-level factorization circuit. (a) Surface-code–compatible logical circuit derived from Fig.~\ref{fig:illustration}a using the logical building-block gates of Fig.~\ref{fig:logical_to_surface}b.
(b) Surface-level implementation of the factorization circuit with 18 surfaces. The computational surfaces \(q_1\)–\(q_5\) (colored) are continuously teleported to satisfy the logical gate implementations geometric constraints. Fault-tolerant (FT) gates are indicated in blue–purple, and magic-state (MS) initializations in green. Because each logical gate spans multiple surface-level timestamps, we parallelize their execution when possible. Insets show top views of timestamps 7 and 18, corresponding to parts of the gates highlighted in (a). Our manual layout could be further optimized for surface count, circuit depth, or feed-forward timestamps between the conditional gates and their control measurement. Feed-forward can be delayed when commuting with subsequent gates (timestamps 20–31 on \(q_0\), surface \((2,2)\)) or be delayed until an ancilla becomes available (timestamps 12–15 on \(q_3\), surface \((1,1)\)).}}
\end{figure*}

The first step is to convert the original circuit from Fig.~\ref{fig:illustration}a into a circuit compatible with surface codes, composed from the logical-gate building blocks shown in Fig.~\ref{fig:logical_to_surface}b. In our circuit, we modify each Toffoli gate of the original circuit into seven \(T\) gates, magic-state initializations, CNOT gates, and \(S\) or \(X^{1/2}\) gates~\cite{amy2013meet}, resulting in the circuit shown in Fig.~\ref{fig:surface_level}a. Several methods can be used to optimize the conversion from a general logical circuit to a surface-code-compatible circuit, such as ZX-calculus techniques~\cite{coecke2023basic} or \(T\)-reduction tools~\cite{kissinger2020reducing, de2019techniques}. This step will be crucial for application-level circuits, where every logical gate introduce a significant overhead\cite{gidney2025factor}. 

 The circuit from Fig.~\ref{fig:surface_level}a is neither unique nor optimized with respect to any parameter. An equivalent circuit can be use fewer logical qubits in expense of a longer logical circuit when using mid-circuit logical measurements~\cite{beauregard2002circuit} (which, in QEC, will result in another decoding-dependent operation). Another potential reduction can possibly be achieved by removing the logical qubit \(q1\), since it is only involved in Clifford gates. We keep this circuit un-optimized, since the goal of our manuscript is to derive the CDS requirement for a mid-range QEC circuit, of which we suggest the circuit in Fig.~\ref{fig:surface_level}a, comprising 14 \(T\)-gates, as one potential candidate.

\subsection{The surface-level circuit}

Given a QEC-compatible circuit, the next challenge is to compile it into a circuit using the surface-level implementations in Fig.~\ref{fig:logical_to_surface}b, while taking into account the physical constraints. We consider the case where the physical qubits are fixed in a two-dimensional plane and have nearest-neighbor connectivity, similar to recent demonstrations of QEC with superconducting qubits~\cite{google2025quantum}. We assume flexibility in the qubit allocation, in the sense that we can freely add ancillary qubits and surfaces as needed in order to run the algorithm, envisioning a dedicated superconducting chip for a target algorithm. 

The physical-qubit layout constraints the surfaces to remain arranged in a grid (as in Fig.~\ref{fig:logical_to_surface}c). Each grid coordinate corresponds to a distance-$d$ surface that includes $d^2 - 1$ physical qubits (in the rotated-surface implementation). On each surface in our implementation, the left and right edges allow an $XX$ surgery and the top and bottom edges allow a $ZZ$ surgery. This imposes a constraint on the locations of the logical qubits, which must be connected by a physical path with the correct logical $Z$ or $X$ edges in order to implement a CNOT gate. Naively, an $N\times N$ surface array can be used to execute an $N$-qubit Clifford algorithm, with the computational qubits placed along the diagonal and paths of ancillary surfaces connecting each pair of qubits which should be added on top of the ancillary surfaces of the non-Clifford gates. Finding an optimal allocation is assumed to be NP-hard~\cite{herr2017optimization}. Mapping schemes are currently under active development~\cite{lao2018mapping, beverland2022surface, akahoshi2024partially}, and recent work shows linearly scaling layouts in partially-FT algorithms~\cite{akahoshi2024partially}.

The complete surface-level factorization circuit, expressed in native surface-code gates, is depicted in Fig.~\ref{fig:surface_level}(b) and uses 18 surfaces. The five computational qubits, highlighted in the figure, are initialized during the circuit and remain active, either by idling or via teleportation, until their logical measurement. Thirteen ancillary surfaces are used to implement the quantum logic under the constraints imposed by the logical $Z$ or $X$ edges of each logical qubit and by the surface-level gate sequence. These additional surfaces are necessary for the logical \(T\), \(S\), and CNOT gates. A video of the resulting physical-level gate sequences for this circuit (surface distance $d=3$) is available in Ref.~\cite{Kurman2025video}.

The inset of Fig.~\ref{fig:surface_level}b shows a top view of the surface grid at timestamps 7 and 18. The top views denote the location of the logical qubits, with diagonal connections between them that allow CNOT operations. The figure illustrates how the primitive surface-level gates belonging to several logical gates are parallelized (encircled in Fig.~\ref{fig:surface_level}a). For example, timestamp 7 includes three parallel $XX$ surgeries corresponding to two logical CNOTs and the teleportation of $q0$, a magic-state initialization towards a $T^{\dagger}$ gate, and an ancillary-surface measurement that finalizes a $T$ gate. 

With this parallelization, the layout includes 40 surface-level timestamps, each implemented physically with \(d\) stabilizer rounds (as shown in Fig.~\ref{fig:logical_to_surface}). As explained in Sec.~\ref{sec:surface_gates}, a single surface-level timestamp is sufficient to implement the FT gates (single-qubit initializations and measurements, and lattice surgeries). A single surface-level timestamp is also shown as the footprint of magic-state initializations, corresponding to non-FT magic-state injection, with or without post-selection. Additional footprint in timestamps and surfaces is required when using magic-state initializations based on distillation or cultivation. This additional footprint is not included in our analysis, as the injection error is assumed to be low enough to provide a reliable output to the target circuit (shown below). However, in large-scale algorithms, this extra footprint, on top of the resources captured in our analysis, may dominate the overall algorithm time or space~\cite{gidney2025factor}

Apart from the magic-state initialization, non-Clifford gates also include decoding-dependent feed-forward operations, colored in orange in Fig.~\ref{fig:surface_level}(b). The number of timestamps available for the CDS to process the QEC physical measurements and apply the feed-forward, without delaying the circuit, corresponds to the horizontal length of each orange line. The feed-forward latency demands are relaxed in our circuit by (i) propagating the feed-forward operation through the circuit until the next non-commuting gate (timestamps 20–31, surface \((2,2)\)), and (ii) by exploiting idling periods when logical surfaces wait for an available ancillary qubit (timestamps 12–15, surface \((1,1)\)). We further discuss the decoding aspects of the circuit in Sec.~\ref{sec:decoding}. 

Table~\ref{tab:surface-summary} summarizes the parameters of the surface-level implementation of the factorization circuit. Some of these parameters may be reduced under specific optimizations, for example by space–time trade-offs. Reducing the number of surfaces is clearly possible at the expense of increased circuit depth, since the surfaces at coordinates \((3,4)\) and \((0,1)\) are used for only three timestamps. In addition, the feed-forward latency demands may be eased by using alternative $T$-gate implementations, such as the ``$\pi/4$-autocorrected'' scheme~\cite{litinski2019game}, at the expense of additional surfaces, total circuit time, and FT Clifford gates. Overall, the requirements that we derive in the next sections are relevant to all non-Clifford surface-code circuits with similar parameters as summarized in Table~\ref{tab:surface-summary}.

\begin{table}[t]
  \centering
  \caption{Summary of the surface-level circuit parameters according to the implementation in Fig.~\ref{fig:surface_level}b}
  \label{tab:surface-summary}
  \begin{tabular}{l r}
    \hline
    \textbf{Parameter} & \textbf{Size} \\
    \hline
    Total surfaces used & 18 \\
    Average number of active surfaces & 7.5 \\
    Total measurements & 105 \\
    Feed-forward gates & 13 \\
    Average fault-tolerant gates (with idling) & 296 \\
    Average magic-state initializations & 13.5 \\
    Average feed-forward latency [\(d\) rounds] & 2.1 \\
    \hline
  \end{tabular}
\end{table}

\subsection{Surface-level circuit error estimates}
\label{sec:eror_estimate}
In this section, we estimate the logical circuit error given the surface-level circuit, and compare it to the error tolerated by the algorithm. This method allows to estimate the required physical error rates, the code distance, and the magic-state initialization scheme. That is, we seek physical-level parameters that reach a surface-level circuit error which remains below the maximal value compatible with a reliable (algorithm-dependent) output probability distribution. 

The circuit's error budget is estimated by categorizing the surface-level gates into FT gates and magic-state initializations, counting the number of gates of each type within the circuit (see Table~\ref{tab:surface-summary}), and calculating their individual error contributions. This estimate is especially useful when the variation between surfaces is small (i.e., their physical qubits follow a similar error model). In that case, instead of simulating the entire circuit, a QEC simulation of a single FT gate and a single magic-state initialization can already provide valuable guidance for the circuit design. If the error rates vary significantly between surfaces, the same estimation method can still be applied by summing the contributions for each surface according to its assigned gates multiplied by its corresponding error parameters.

For our target factorization circuit, we show in Appendix~\ref{app:circuit} that the ideal output distribution on the logical qubits $q0$–$q2$ has three high-probability outcomes ($\ge 0.235$), while all remaining outcomes have probabilities $\le 0.063$. For this distribution, a maximal circuit error of $8.6\%$ (in total-variation distance) is sufficient to keep the three target outcomes statistically dominant over all others (see Appendix~\ref{app:circuit} for details). We therefore seek physical error rates, code distance, and a magic-state initialization scheme that keep the overall errors below a $8.6\%$ threshold.

\begin{figure} [t]
\includegraphics[width=\columnwidth]{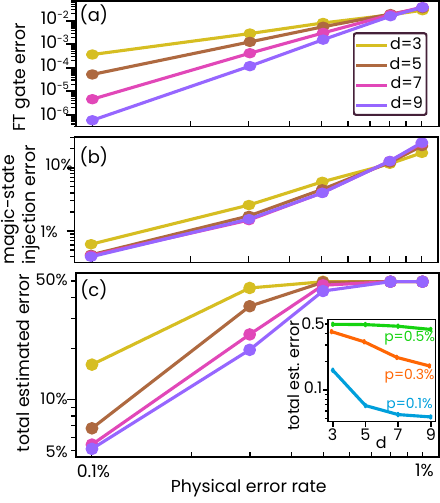}
 \caption{\label{fig:errors} Error estimation for the factorization circuit.
(a) Single FT error example: logical error from idling for $d$ stabilizer rounds.
(b) Error from a non-FT magic-state initialization using state injection. Increasing code distance does not suppress this contribution.
(c) Estimated total error for the circuit, computed binomially as the probability of an odd number of flips across 14 nFT gates and 296 FT gates. Inset: total error versus distance $d$, saturating near $\sim5\%$. From this estimate, achieving a circuit error below 8.6\%, as needed from our circuit (see Appendix~\ref{app:circuit}) requires a physical error rate of $\approx0.1\%$ and $d\!\ge\!5$.}
\end{figure}

As explained in Sec.~\ref{sec:Clifford}, the FT gate set consists of single surface-level timestamps of idling, Pauli-state initializations or measurements, and parity measurements in the \(Z/X\) bases. FT gates exhibit exponential error reduction when increasing distance below the error threshold (after decoding), and their error rates are determined by the logical error of \(d\) stabilizer rounds on a single surface, i.e., a volume of order \(d^3\). Lattice surgery is counted as two FT gates, since it involves two surfaces. The \(\ket{S}\) initializations or measurements in the \(Y\) basis are also counted as two FT gates~\cite{gidney2024inplace}. Figure~\ref{fig:errors}a shows the FT error, plotting the simulated idling error for a distance-\(d\) surface over \(d\) stabilizer rounds.

Different manifestations are possible for initializing magic-states in surface-codes, with significant tradeoffs (further explained in Sec.~\ref{sec:non-Clifford}). The goal here is to find the simplest initialization method that can support a circuit error within the error budget defined by the algorithm. For our target circuit, we suggest using a non-FT magic-state initialization, using the hook-injection technique~\cite{gidney2023cleaner}) and post-selection. This technique allows initialization of a distance-\(d\) surface in \(d\) stabilizer rounds (a single surface-level timestamp) without additional qubits. A post-selection (or repeat-until-success in real implementations) keeps only the shots which did not include any syndrome event, i.e., when all stabilizer measurements during the injection were measured as '0'. Figure~\ref{fig:errors}b presents the injection error simulation results, showing how the initialization error are distance agnostic (beyond distance-$3$).  
  
Figure~\ref{fig:errors}c presents the total estimated error of the factorization circuit, obtained by combining the error outcomes of the two small-scale simulations in Figs.~\ref{fig:errors}a,b. We find that a physical error rate of \(P_{\mathrm{phys}} = 0.1\%\) with \(d \ge 5\) is sufficient to reach a total error below the \(8.6\%\) threshold, whereas for \(P_{\mathrm{phys}} = 0.3\%\) the total error remains above this bound for all distances considered. This estimate shows that, for \(P_{\mathrm{phys}} = 0.1\%\) and \(d \ge 5\), the magic-state initializations through injection dominate the error budget, leading to a saturation of the total fidelity as the distance is increased (see the inset of Fig.~\ref{fig:errors}c). 

The error budget can therefore be interpreted as follows. At fixed \(P_{\mathrm{phys}}\), the FT contributions decrease with increasing \(d\), while the non-FT magic-state initialization contribution is essentially distance-independent. In the regime where the total error has saturated, any residual differences between curves at different distances originate from the FT part of the circuit, whereas the used magic-state initializations provide an distance-independent error. 

For larger circuits or more stringent logical error requirements, this picture suggests two main levers: (i) adopting higher-fidelity magic-state initialization schemes (e.g., distillation or cultivation) to reduce the magic-state initialization contribution, and (ii) increasing the code distance \(d\) to further suppress the FT contribution. In the future, we believe that large-scale non-Clifford algorithms will often require both measures, together with circuit-level optimizations that reduce the overall gate count. This back-of-the-envelope estimate can be useful for logical or surface-level circuit optimization when including an error variations between surface-coordinate.
 
These simulations, and the large-scale ones (in Sec.~\ref{sec:phys_sim}), uses \texttt{stim} simulator \cite{gidney2021stim}. We use an error rate of \(p_{\mathrm{phys}}\) for depolarizing any two-qubit gate and flipping a physical measurement, while the single-qubit gate error was \(p_{\mathrm{phys}}/10\). This circuit-level model reflects the empirical hierarchy on current superconducting devices \cite{google2025quantum}, and is a useful single-parameter abstraction for qualitative QEC studies. On real hardware, noise is often biased and correlated (e.g., crosstalk, leakage), which practical decoders should cover to maintain a high error threshold.

\section{Physical-level simulations}
\label{sec:phys_sim}

After reaching the surface-level circuit and determining the magic-state initialization method expected, we now turn to the circuit's end-to-end physical-level implementation, involving thousands of qubits. These simulations establish the surface-level to physical-level compilation, verify the required physical-level error rates, and are useful for deriving the physical-level control requirements of QEC circuits of similar magnitude. To execute correctly these physical-level simulations, we constructed additional functionalities on top of the commonly used  \texttt{stim} Python package~\cite{gidney2021stim}, that enable coding of the surface-level circuit in a logical \& surface-level intermediate representation from which the complete physical-level sequences, errors, and decoding graphs, are automatically compiled (see Appendix~\ref{app:sim}).

The advantage of using \texttt{stim} is that it enables the simulation of noisy quantum circuits at the scale of thousands of qubits including the mapping of the physical measurement data into a weighted decoding graph according to an input error model, and integrates with decoders such as \texttt{pymatching}~\cite{higgott2022pymatching}. However, the ability of \texttt{stim} to simulate noisy circuits in the thousand-qubit scale relies on restricting the dynamics to the Clifford group. As a result, it cannot include any non-Clifford gate, and thus the magic-state initialization. In addition, \texttt{stim} (at least in the version which we have used) does not incorporate decoding-dependent feed-forward (mid-circuit) gate modifications. These are key features of any non-Clifford circuit. 

Our approach to overcome these limitations is as follows. The errors and performance of the magic-state initialization are inferred by simulating the initialization of a Clifford state, such as an \(\ket{S}\) state, using the same gate sequence that would be used for preparing a \(\ket{T}\) state, up to few single-qubit non-Clifford gates for each surface initialization (which are now Clifford gates). This methodology is commonly used to analyze the performance of magic-state initialization schemes using \texttt{stim}~\cite{gidney2023cleaner,gidney2024magic}. The decoding-dependent feed-forward gates are more challenging to capture within \texttt{stim}. We simulate the circuit without these operations, and extract the additional error due to extended decoding times separately  (in Sec.~\ref{sec:decoding} and Fig.~\ref{fig:decoding_error}). Thus, although simulated circuit is not the exact physical-level circuit which implements the surface-level circuit in Fig.~\ref{fig:surface_level}b, it represents the expected errors from the circuit's physical implementation up to the decoding latency delay errors which are analyzed and added separately.

\begin{figure*} [t]
\includegraphics[width=\textwidth]{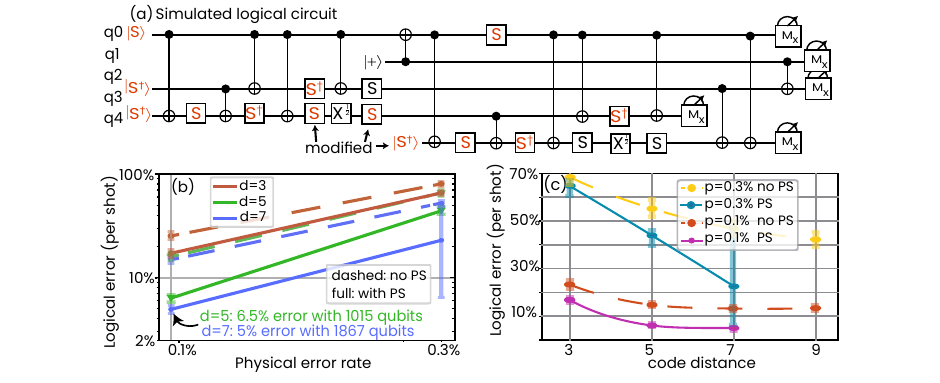}
 \caption{\label{fig:sim} Physical-level simulation results of the factorization circuit.
(a) Simulated logical circuit mirroring Fig.~3(a), with $T$ gates replaced by $S$ gates implemented (marked in red). The simulated physical-level implementation omits feed-forward.
(b) Logical error versus physical error, with and without post-selection (PS). Because of non-FT magic-state initializations, the gain at $d{=}7$ is negligible relative to $d{=}5$ for physical errors of $0.1\%$. Target logical error of $8.6\%$ is out of reach at $0.3\%$ physical error. Error bars indicate $99.9\%$ binomial confidence intervals.
(c) Logical error versus code distance $d$, showing saturation with increasing $d$ due to dominance of magic-state initializations errors.}
\end{figure*}

Following these modifications, the simulated logical circuit is depicted in Fig.~\ref{fig:sim}a. We replaced the logical \(\ket{T}\)-state initializations with \(\ket{S}\)-state initializations and all \(T\) gates with \(S\) gates. At the surface level, these modifications correspond to replacing the magic-state initializations (green in Fig.~\ref{fig:surface_level}b) with \(\ket{S}\)-state initializations. At the physical level, where we implement the magic-state initialization with state-injection \cite{gidney2023cleaner}, the change is a single physical-qubit gate per initialization, altering a \(X^{1/4}\) rotation with a \(X^{1/2}\) rotation, while leaving the rest of the initialization gate sequence unchanged. With this change, we can verify the correctness of the logic, without any further circuit modifications, through 3 out of 5 logical stabilizers, \(X_{q_0}X_{q_4}\), \(X_{q_1}X_{q_2}\), and \(X_{q_3}\), which we define as the simulation observables for the circuit. We do not check the remaining two stabilizers as they correspond to logical measurements that require additional circuit modifications in the form of logical S rotations. Each of the checked observables is associated with decoding graphs containing up to tens of thousands of nodes, and any error in the physical gate sequence or in the mapping of physical measurements to graph nodes would prevent the decoding graph from compiling and would preclude obtaining meaningful (non-random) logical-level decoding outputs.

A detailed description of the simulation structure is provided in Appendix~\ref{app:sim} and Fig.~\ref{fig:sim_SM}. Briefly, as in Fig.~\ref{fig:surface_level}, each surface is defined by its coordinate in a two-dimensional grid and its orientation (which edges correspond to the logical \(Z\) and \(X\) operators). The surface-level operations in each timestamp are translated to their physical implementation, consisting of \(d\) stabilizer rounds. All gates in the \(Z/X\) bases, including surface initializations, measurements, and lattice surgeries, are implemented fault-tolerantly at the physical level (as explained in Sec.~\ref{sec:Clifford}), while \(\ket{S}\)-states (or \(\ket{\sqrt{X}} = \ket{+} + i\ket{-}\)) are prepared using hook injection~\cite{gidney2023cleaner}. The full video of the physical-level gate sequences for distance \(d=3\) is presented in Ref.~\cite{Kurman2025video}.

\begin{table*}[t]
  \centering
  \small
  \setlength{\tabcolsep}{4pt}
  \caption{Physical-level implementation aspects, as extracted from simulations.}
  \label{tab:phys-resources}
  \begin{tabular}{l c c c c c c c}
    \toprule
    & No.\ of physical qubits
    & Total stabilizer rounds
    & max active qubits
    & max parallel 2-Q gate
    & max parallel measurements
    & No.\ of physical measurements
    & Avg data creation [bit per QEC round] \\
    \midrule
    scaling & $d^2$ & $d$ & $d^2$ & $d^2$ & $d^2$ & $d^3$ & $d^2$ \\
    $d=3$ & 419 & 117 & 232  & 88   & 112  & 8061   & 69   \\
    $d=5$ & 1015 & 195 & 632  & 268  & 312  & 36687  & 188  \\
    $d=7$ & 1867 & 273 & 1224 & 544  & 608  & 96006  & 363  \\
    $d=9$ & 2975 & 351 & 2008 & 916  & 1000 & 203874 & 595  \\
    \bottomrule
  \end{tabular}
\end{table*}

Importantly, we retain part of the original logical-level circuit information when constructing the decoding graphs which affect the logical circuit's results. Knowing the order of the non-commuting logical gates and their surface-level implementations is essential for connecting the logical circuit with the Pauli frame corrections which arise from lattice-surgeries or ancillary qubit measurements (discussed in Sec.~\ref{sec:surface_gates}). These dependencies are compiled using our intermediate-level representation when describing each surface-level gate not only by its participating surfaces and surface-level timestamp, but also by the logical gate which is being implemented and its position within the logical circuit. 

Because each logical gate is realized by several surface-level operations, their execution order in the surface-level schedule does not necessarily coincide with their order in the logical circuit. Therefore, we add a compilation layer which constructs the decoding graphs according to the XOR (exclusive-or) of all logical surface measurements which might flip the final logical operator of interest. These decoding graphs eventually include all physical measurements within all stabilizer rounds and data qubit measurements which may flip the logical outcomes. This procedure relief the need of any unnecessary Pauli feed-forward physical correction to the quantum circuit. \texttt{stim}'s compiler verifies that the decoding graphs are consistent and there is no error in their construction. Overall, our simulation structure allows the execution of QEC algorithms while treating the code distance and error-model parameters as free inputs.

Figures~\ref{fig:sim}b and \ref{fig:sim}c present the simulation outputs: the logical errors of the physical implementation of the surface-level circuit as a function of code distance, for physical error rates of $0.1\%$ and $0.3\%$, with and without post-selection (PS) during the magic-state injection. These plots indicate that both a low physical error rate of $0.1\%$ and post-selection during magic-state injection are necessary to achieve sufficient circuit fidelity. At a physical error rate of $0.3\%$, none of the examined code distances reached the target error threshold, regardless of post-selection. At $0.1\%$, logical errors were dominated by magic-state initialization, with increasing code distance having only a minor effect. Due to this error saturation with distance, a surface distance of $d=5$ was sufficient to reduce the circuit error below $8.6\%$. These requirements on code distance and physical error rate, as well as the observed saturation of logical error with distance, are consistent with the anticipated error-summation behavior shown in Fig.~\ref{fig:errors}c. The exact simulation errors differ slightly from those in Fig.~\ref{fig:errors} due to the inclusion of additional qubits associated with lattice surgery.

From this analysis we conclude that approximately $1000$ physical qubits are sufficient to execute QEC non-Clifford circuits containing $\sim 15$ non-Clifford gates, with a circuit target error above $7\%$ when the physical error rate is $0.1\%$. Achieving lower circuit errors requires a higher-fidelity magic-state initialization scheme. Only then should increasing the code distance be considered. These findings emphasize the implications for CDS design: maintaining a sufficiently low physical error rate $P_{\mathrm{phys}}$ is essential. High-fidelity analog control, accurate readout, and low-noise electronics are required to keep $P_{\mathrm{phys}}$ in the $\sim 10^{-3}$ regime or below. Otherwise, even large code distances and advanced magic-state protocols will be insufficient to achieve the desired circuit error.

In addition to characterizing logical error behavior, the physical-level simulation enables extraction of resource estimates for the factorization circuit, as summarized in Table~\ref{tab:phys-resources}. These estimates are particularly relevant for future experimental demonstrations of QEC circuits with comparable numbers of logical qubits and surface-level gates and without magic-state distillation or cultivation overhead. Regarding qubit count, the required quantum chip size ranges from several hundred to a few thousand physical qubits, scaling as $d^2$. For circuit run-time, executing a single shot of the complete factorization circuit requires approximately $100$--$350~\mu\text{s}$ on superconducting qubits, since the total number of stabilizer rounds scales only linearly with $d$.

For the controller part of the CDS design, a critical resource metric is the number of simultaneous control channels needed to implement physical gates. Although single-qubit gates must be applied to all active qubits in parallel, implying that the number of single-qubit control channels should equal the total qubit count, our implementation shows that only up to $55$--$65\%$ of the qubits are active simultaneously (column~3). This reduction enables multiplexed control, where the same control lines are shared among multiple qubits, with signals switched between them at the cryogenic stage during circuit execution. A similar observation holds for two-qubit gates and readout (columns~4 and 5). The maximum parallel readout further defines the requirements for classical analog-to-digital signal processing throughput, as the control system must convert up to thousands of analog measurements into digital data. The average measurement rate (column~6), which corresponds to the data-creation rate, is a critical determinant of data transfer between the controller and decoder in a CDS, setting the required channel bandwidth in bits per second. For superconducting qubits with stabilizer rounds of $1~\mu\text{s}$ duration~\cite{google2023suppressing}, bandwidths of $1\,\text{Mbit/s}\times\text{qubits}$ are sufficient to transfer all measurement data. Finally, the total number of measurements provides a scaling estimate for the data volume the decoder must process, which grows as $d^3$.

\section{Decoding requirements for near-term non-Clifford circuits}
\label{sec:decoding}

\begin{figure*} [!t]
\includegraphics[width=\textwidth]{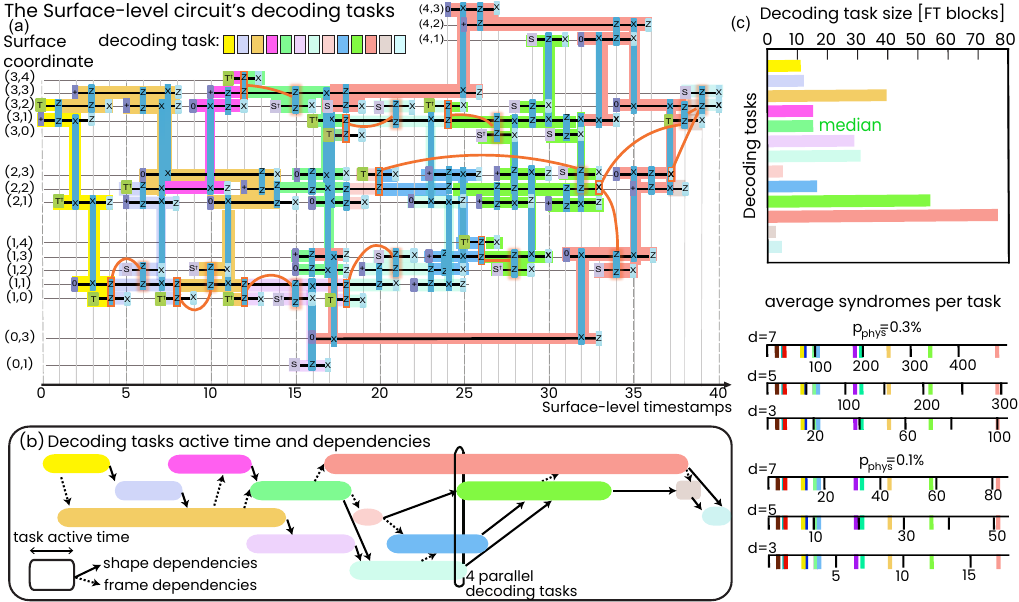}
 \caption{\label{fig:decoding} Decoding aspects of shallow non-Clifford circuits.
(a) Surface-level circuit from Fig.~\ref{fig:surface_level}b showing 13 color-coded decoding tasks.
(b) Active time windows for each task and their inter-task dependencies (arrows). In our surface-level implementation, the decoder must process at least five tasks simultaneously. Interdependencies include shape (solid arrows) and frame (dashed arrows) dependencies, where one decoding output modifies the shape or boundary-condition Pauli frames of another, respectively.
(c) Decoding task size measured in FT $d^3$ blocks (top) and average number of syndromes per task (bottom), shown for different code distances $d$ and physical error rates $p_{\text{phys}}$. Separate axes are provided for each $d$ and $p_{\text{phys}}$.}
\end{figure*}

The decoder is a classical processor that detects local quantum errors based on physical QEC measurement results and an error model for the gate sequence executed by the controller. The field of decoding algorithms is rapidly advancing, with solutions exhibiting trade-offs among accuracy, decoding time, and classical resources~\cite{delfosse2021almost,wu2023fusion,battistel2023real,kolmogorov2009blossom,chubb2021general,das2022lilliput,meinerz2022scalable,higgott2025sparse}. As discussed in Sec.~\ref{sec:non-Clifford}, achieving quantum computation with QEC relies on the CDS’s ability to apply mid-circuit, decoding-dependent gates for every non-Clifford operation. Recent studies have shown that control-system latency and tight controller–decoder integration are critical for the successful implementation of large-scale non-Clifford circuits~\cite{maurya2024managing,kurman2025benchmarking}, motivating a focus on CDS architectures designed for low-latency feed-forward. In this chapter, we examine key aspects of real-time decoding requirements, including the division of the physical circuit into decoding tasks, specific CDS decoding latency values, and the influence of code distance and physical error rates, based on our proposed near-term use case.

In our example, decoding the logical measurement value of the lattice surgery between a logical qubit and a T-state determines whether another lattice surgery between the same logical qubit and an S-state should be performed (i.e., the decoding-dependent gate). These decision-making events define the division of the surface-level circuit into 13 distinct decoding tasks, shown in Fig.~\ref{fig:decoding}a, where each highlighted color represents a separate task. Each decoding task terminates at the end of the lattice surgery with a T-state (or at the end of the circuit) and begins either when another task ends or at the start of the circuit, ensuring no overlap between tasks to minimize double decoding of the same syndromes. Importantly, the boundary between two decoding tasks is not strict, as certain errors may only be fully detectable only few QEC rounds after the lattice surgery has concluded. Our task-division method minimizes the number of inter-task boundaries and multiple decoding of the same syndromes; however, alternative divisions are possible—for example, assigning each task to decode from the beginning of the circuit or employing sub-tasks and fusion, as in the fusion-Blossom algorithm~\cite{wu2023fusion}. The critical requirement is that the decoding unit produces the task’s logical output and communicates it to the controller and the dependent tasks as quickly as possible.

Figure~\ref{fig:decoding}b illustrates the active duration of each task and their interdependencies. We find that non-Clifford circuits require the decoding unit to execute multiple decoding tasks in parallel, up to four in this example, and to enable communication between tasks for updating Pauli frames and task shapes. Two types of dependencies arise. First, feed-forward gates introduce a shape dependency between a decoding task and its predecessors (indicated by solid arrows): if the decoding task output delays beyond the expected feed-forward time or the lattice-surgery output of a preceding task is ``1'' (in our T-gate implementation), the subsequent decoding task becomes larger~\cite{kurman2025benchmarking}. That is, the number of physical measurements to analyze in a decoding task (i.e., its size) depends on the prior decoding outcome and its latency. Second, a frame update occurs when two decoding tasks share a boundary or when a logical measurement in one task can flip the output of another. In our \(T\)-gate implementation, shape dependencies must be communicated immediately between decoders. In contrast, other implementations~\cite{litinski2019game} can convert some of the shape dependencies into frame dependencies, which only need to be transferred before the feed-forward gate of the receiving task.

This analysis of a full non-Clifford circuit highlights the need for a new perspective on the decoder architecture within the CDS: shifting from a single decoder handling the entire QEC experiment to a decoding unit that processes multiple tasks in parallel, with explicit dependencies and communication channels between them. This architecture can be implemented by allocating sets of CPU or GPU cores to each decoding task, where tasks support shape and frame dependency managements. An implementation using FPGA decoders is also suitable for the multi-task decoder when including a coordinator unit~\cite{wu2024lego}.

To estimate the classical processing load requirements on the decoding unit, we calculate the expected number of detected syndromes (either physical or measurement errors) within each decoding task. This estimation is based on the space–time volume of the task, defined as the number of FT blocks (\(d\) stabilizer rounds for a distance-\(d\) surface). Figure~\ref{fig:decoding}c shows the space–time volume of each decoding task in the factorization circuit (top) and the corresponding average number of syndromes for different distances and physical error rates (bottom). For low error rates (\(0.1\%\)), we find that a significant portion of tasks contain only a few tens of syndromes on average. These low numbers motivate the development of embedded decoders within the controller (e.g., FPGA-based decoding schemes~\cite{liyanage2023scalable}), as suggested in various micro-architecture proposals~\cite{das2022afs} for similar near- to medium-term demonstrations. Additionally, a pre-decoder stage, as recently proposed~\cite{smith2023local, caune2023belief}, could resolve a substantial fraction of syndromes when they are sparse in the decoding graph, leaving only the remainder to be handled by fast, dedicated decoding hardware (e.g.,~\cite{barber2025real, das2022afs, bausch2024learning}).

Another key question in CDS designs is the impact of the controller–decoder communication round-trip time and the decoding latency. We combine these into a single parameter, CDS latency, defined as the time interval between the last measurement of a decoding task and the execution of the first mid-circuit gate that depends on it. This parameter is critical because QEC rounds, implemented as logical idling, must continuously run on the surfaces of the logical qubits to preserve the encoded quantum state. Consequently, the size of the pending task will continue to grow beyond the values shown in Fig.~\ref{fig:decoding}c until the decoding output is received. At scale, this size dependency is the root cause of potential catastrophic backlogs~\cite{terhal2015quantum}, creating a requirement for sub-linear decoder complexity so that the decoder's throughput exceeds the syndrome data generation rate~\cite{kurman2025benchmarking}.

\begin{figure} [t]
\includegraphics[width=\columnwidth]{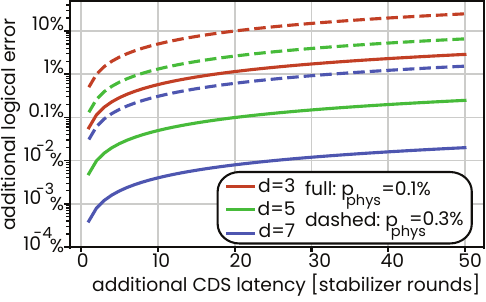}
 \caption{\label{fig:decoding_error} Added circuit error vs the additional CDS latency which delays the circuit until the feed-forward is applied. Larger $d$ and lower $p_{\text{phys}}$ relax latency requirements. For example, every 20 rounds of delay at $p_{\text{phys}}=0.1\%$ and $d=5$ adds only $\sim0.1\%$ to the total circuit error.}
\end{figure}

In our surface-level circuit design, most feed-forward gates are not placed immediately after their corresponding task. This time gap creates a task-specific latency budget: if the CDS latency remains below this budget, decoding and communication delays do not alter the quantum circuit or introduce additional errors (the quantum-limited regime in Ref.~\cite{kurman2025benchmarking}). If the CDS latency exceeds this budget, at least one logical qubit must idle, increasing the overall logical error of the circuit. An upper bound on the additional logical error due to feed-forward latency for different distances and physical error rates is shown in Fig.~\ref{fig:decoding_error}. This error is calculated as
\begin{equation}
  P_{\mathrm{delay}}
  = N \cdot P_{\mathrm{FT}} \cdot \frac{T_{\mathrm{delay}}}{d \cdot T_{\mathrm{round}}},
  \label{eq:pdelay}
\end{equation}
where \(T_{\mathrm{delay}}\) is the CDS additional latency beyond the budget of each task, \(T_{\mathrm{round}}\) is the duration of a single stabilizer round, and \(P_{\mathrm{FT}}\) is the error of a single FT gate per \(d\) rounds for one surface. \(N\) denotes the number of logical qubits in the circuit (five here), which must remain idle until the feed-forward is applied. We assume no classical errors occur during controller–decoder communications, given the negligible error rates of expected protocols (e.g., RDMA over Converged Ethernet).

For the superconducting-qubit implementation of our circuit, if the total \(T_{\mathrm{delay}}\) across all decoding tasks remains below a few tens of \(\mu\text{s}\), the additional error is negligible compared to the rest of the quantum circuit. As shown above, without considering this delay, achieving a logical error below the target of 8.6\% requires a physical error rate of 0.1\% and a code distance of at least 5. Within this regime, even a 50~\(\mu\text{s}\) delay adds less than 0.3\% (or 0.02\%) error for distance-5 (or distance-7) circuits. The maximum total delay budget for a distance-5 circuit is on the order of 400~\(\mu\text{s}\), determined by the additional error that closes the gap between the simulated circuit without delay and the target error. We note that increasing the distance improves tolerance to delay through two factors: the denominator in Eq.~\eqref{eq:pdelay} and a smaller \(P_{\mathrm{FT}}\).

The analysis presented above applies to any decoding-dependent protocol, though the allowed delay depends on the circuit's target logical error. For example, the qubit-efficient 15-to-1 magic-state distillation scheme~\cite{litinski2019game} involves five logical qubits and 11 decoding tasks, requiring the decoder to be organized into multiple units and necessitating communication between them. The permissible circuit delay, derived from the additional error introduced by CDS latency, is expected to follow Eq.~\eqref{eq:pdelay}. For concrete values, the scheme allows \(N = 1\) for each task, such that maintaining \(P_{\text{delay}} < 10^{-8}\) for a distance-15 code (where \(P_{\text{FT}} \approx 5 \times 10^{-10}\) for \(P_{\text{phys}} = 0.1\%\)) requires a total additional CDS latency below \(20d\) stabilizer rounds. This corresponds to approximately \(300~\mu\text{s}\) in total, translating to an average extra latency of about \(27~\mu\text{s}\) per task, on top of the \(15~\mu\text{s}\) latency per task permitted by the protocol. This latency is relaxed in magic-state cultivation, which involves a single decoding task per magic-state initialization, such that a single-task CDS latency of \(300~\mu\text{s}\) can support a similar target of \(P_{\text{delay}} < 10^{-8}\). Thus, in terms of decoding-dependent latency, the cultivation scheme appears to offer an order-of-magnitude relaxation in CDS latency requirements for keeping high-fidelity magic-state initialization.

\section{Summary and Discussion}

Throughout this manuscript, we have examined key aspects of the control system required for the successful execution of a non-Clifford circuit with surface QEC code. Below, we summarize our main findings, grouping the requirements into three categories: the controller, the communication channel, and the decoding unit.

\begin{itemize}

    \item \textbf{Controller}
    \begin{itemize}
        \item Low coherent control errors (\(\sim 0.1\%\))
        \item QPU–controller latency of 100 ns
        \item Decoding-dependent pulse branching (conditional commands, while loops)
    \end{itemize}

    \item \textbf{Communication channel}
    \begin{itemize}
        \item Few \(\mu\text{s}\) latency communication protocol
        \item Channel bandwidth \(\sim 1 \,\text{Mbit/sec} \times \text{qubits}\)
        \item Syndrome allocation within decoders
    \end{itemize}

    \item \textbf{Decoding unit}
    \begin{itemize}
        \item Parallelize decoders
        \item Real-time decoding task modification
        \item Communication between decoding tasks
        \item decoding latency up to few tens of \(\mu\text{s}\) 
    \end{itemize}
\end{itemize}

Regarding the controller, we emphasize the importance of minimizing coherent control errors which are systematic errors arising from imperfect calibrations and control fields. This requirement stems from the need to maintain a physical error rate on the order of \(0.1\%\) for systems with \(\sim 1000\) qubits. A second critical control requirement is ultra-low latency between the controller and the QPU during conditional gate execution for real-time feed-forward operations, which are not necessarily decoding-dependent. This capability underpins real-time code branching, such as in repeat-until-success magic state protocols. Keeping a low QPU--controller latency is also necessary to minimize the controller footprint in the overall CDS decoding latency. Finally, the control flow must support decoding-dependent operations, including conditional statements (\texttt{if/else}) and iterative loops (\texttt{while}), to enable adaptive execution based on decoding-dependent real-time feedback.

Supporting low-latency decoding-dependent feed-forward operations is the primary rationale behind the requirements for the communication channel and the decoding unit. Assuming a worst-case scenario with quantum hardware requiring a microsecond for a stabilizer round, as in superconducting transmon qubits \cite{google2023suppressing}, the total two-way communication and decoding latency must remain within tens of microseconds. This time budget should be allocated primarily to the decoder to maximize decoding time, leaving the communication channel with a latency timescale of a few microseconds. The communication channel should support a bandwidth of around \(\text{Mbit/s}\) for each active qubit if all syndromes are transmitted (1 bit per physical qubit per stabilizer round). However, this requirement can be reduced with a pre-processing stage within the controller to transmit only the detected syndromes.

Maintaining low latency necessitates that the decoding unit can execute multiple decoding tasks in parallel, support dependencies and communication between tasks, and dynamically modify the shape of the decoding tasks in real time. To quantify decoder resources, each decoding task can be associated with a sparse Tanner graph (or the detector matching graph equivalently for MWPM-based decoders) whose size scales with the task space--time volume. For task volumes of up to tens of FT $d^3$ blocks, as shown above, the required hardware memory can reach the MB scale per graph. This follows from the number of check (detector) nodes, which scales as $\mathcal{O}(B\,d^3)$ for a task of size $B$ FT blocks, together with a linear number of edges determined by the graph connectivity, characterized by a constant factor $\kappa$ (typically in the range 1--5 for local noise models).

Although these requirements were derived from a specific circuit, they can serve as general guidelines for any non-Clifford circuit involving approximately 15 \(\pi/4\) logical rotations with a small number of logical qubits and a total error budget of a few percent. Many of these requirements can also be extended to more demanding circuits. Similar compilation stages are expected: first, minimizing the logical circuit (e.g., using ZX calculus \cite{fischbach2025review}, which can reduce gate count by tens of percent); then mapping the logical circuit to a QEC-compatible version based on the building blocks of the chosen code; followed by conversion to the code-level (surface-level in our manuscript) and finally to the physical circuit. All controller requirements, particularly decoding-dependent real-time modifications, remain essential for non-Clifford circuits. Likewise, the decoder must support parallelized tasks that correspond to decision-making events, with dependencies in frames and shapes. At scale, orchestration among multiple decoders will require a dedicated compiler and coordination framework \cite{wu2024lego}. Importantly, these requirements are independent of the specific decoding algorithm or QEC code.

From a hardware perspective, we note that there is no conceptual limit on the number of gates that can be executed in parallel: each parallel gate can be driven from a separate room-temperature control channel or multiplexed within a cryogenic stage. Current control hardware is already mature enough to drive gates through hundreds of synchronized channels in parallel, and channel density is expected to double within the next few years. Practically, however, we do not expect current control solutions for superconducting qubits to support parallel execution of a full \(\sim\)1000-qubit algorithm in the near term, although this may become feasible as channel density continues to scale; comparable parallel control is already achievable with neutral-atom QPUs. Should the achievable hardware parallelism fall short of a circuit's scheduling requirements, gates that cannot be executed concurrently must instead be serialized, increasing the overall circuit execution time and depth. This in turn relaxes the real-time requirement on the decoder, since decoding tasks are correspondingly spread out over a longer time window, at the cost of a longer overall circuit runtime.

As these circuits scale, the number of non-Clifford gates dictates the number of required decoding tasks, and we expect the size of these tasks to grow with the number of logical qubits due to additional connectivity routing. In all such circuits, the total error can be estimated as the sum of errors from FT gates and magic-state initializations (see Sec.~\ref{sec:eror_estimate}). To achieve lower logical circuit errors or increase circuit depth, magic-state initialization cannot rely on the injection method used here. Improvements can be achieved through magic-state cultivation, where the target error is controlled by the level of real-time discarding, code distance, and physical error rate \cite{gidney2024magic}. This approach does not alter any CDS requirements but requires additional space-time volume to prepare states prior to execution. Furthermore, scaling will demand larger code distances to reduce FT errors below magic-state initialization errors. Once this condition is met, requirements on decoding and communication latency will reduce to a 10 $\mu s$ scale to minimize extra FT errors. In any case, CDS decoding latency and communication will not delay the quantum circuit or introduce additional errors, provided the feedback loop remains below \(d\) stabilizer rounds when using the \(\pi/8\) auto-corrected scheme \cite{litinski2019game}.

In parallel, it is crucial to advance computational techniques for quantum LDPC codes~\cite{bravyi2024high}, which rely on long-range interactions. Recent proposals suggest that a 3D architecture could enable computation with a cat-LDPC code~\cite{ruiz2025ldpc}, potentially mitigating the high connectivity overhead of surface codes. In all these scenarios, control requirements are expected to follow similar guidelines, as computation remains measurement-based and demands real-time decoding.

To reach low error rates at scale, scalable qubit characterization techniques~\cite{chen2022calibrated,hockings2025scalable} or error learning~\cite{caune2023belief} are essential, which may be facilitated by the fact that only specific physical gates are required. From an algorithmic perspective, developing a compiler that transforms a logical circuit to a surface-level circuit, considering qubit-topology constraints, would be highly beneficial. Moreover, it is crucial to find ways to decode lattice surgery during circuit execution, enable decoding-dependent mid-circuit operations (a limitation of the simulation within this paper), and enable communication between decoders. 

In conclusion, this paper outlines the key requirements for executing fault-tolerant quantum algorithms, focusing on non-Clifford circuits. By specifying control-system constraints, decoding dependencies, and fidelity considerations, we provide a foundation for implementing these advanced algorithms. As hardware scales to hundreds or thousands of qubits, these insights will guide the CDS development, which are critical for building quantum supercomputers~\cite{mohseni2024build}. Applying these guidelines to real hardware will mark a significant milestone toward scalable, error-corrected quantum computation.

\section*{Author contributions}
Y.K. wrote the manuscript, performed the analysis, and carried out the simulations. L.E. formulated the problem and proposed solution strategies. O.W. contributed to the solution architecture. N.H. developed the simulation framework and code. Y.C. supervised the project as principal investigator. No generative AI or large language model was used in producing this manuscript.

\printbibliography

\onecolumn\newpage
\appendix

\section{Simulation structure}
\label{app:sim}
The simulations presented in this work are available in Ref.~\cite{Kurman2025code}. The simulations are based on the \texttt{stim} Python package~\cite{gidney2021stim}, which supports QEC simulation with thousands of physical qubits, and provides an automatic mapping between syndromes (detectors) and a matching graph. We used decoding via the \texttt{PyMatching} package~\cite{higgott2022pymatching} and \texttt{sinter} for fast QEC Monte Carlo sampling, which we limited to 30{,}000{,}000 shots or 1200 errors. The main challenge in enabling these simulations was converting a logical circuit to the physical implementation of 18 surfaces and 38 surface-level timestamps. It was necessary to keep the exact gate sequence and exact detector definitions to reach a fault-tolerant implementation of the \(Z/X\) logical gates and an \(S\)-state (or \(X^{1/2}\)-state) initialization through magic state injection, while enabling parallel operations across all active qubits.

The simulations were built in two layers. In the bottom layer, we developed the physical-level infrastructure, which specifies (i) the physical structure of a surface or pair of surfaces for lattice-surgery and (ii) the physical implementation of specific surface-level operations under a specified error model. Examples include initializing a surface in a chosen basis, initializing surgery between two surfaces, performing a stabilizer round over all active surfaces, adding logical measurements, and adding observables. Observables are required by the decoder, which outputs if there was a flip in the observable value; such flips occur due to induced physical errors in the simulation. In our simulations, we used a depolarizing error with probability \(p\) for a physical measurement and a two-qubit gate, and \(p/10\) for a single-qubit gate (though the code can support distinct error rates for each physical operation). This physical-level layer is sufficient to generate the required gate sequence for surface-code experiments, but it becomes cumbersome at scale and its connection to the logical circuit is not straightforward.

To simplify implementation and verify the correctness of the surface-level operation sequence, we built an intermediate representation (IR) code. This layer contains information regarding the logical-level circuit (the logical qubits and logical gate sequence, Figure~3a) and the surface-level circuit (surface coordinates and surface-level gate sequence, Figure~3b). This IR layer produces a surface-level quantum circuit that includes observables constructed from all surface-level measurements upon which the observable depends, according to Pauli-frame propagation in the logical circuit. Pauli-frame propagation is crucial for eliminating unnecessary corrective physical gates in the physical sequence. Technically, knowing the logical timestamp and the logical operation associated with each surface-level gate enables correct frame propagation. The resulting surface-level circuit is then verified within \texttt{stim} so that each observable is well defined.

\begin{figure}[b]
\includegraphics[width=\textwidth]{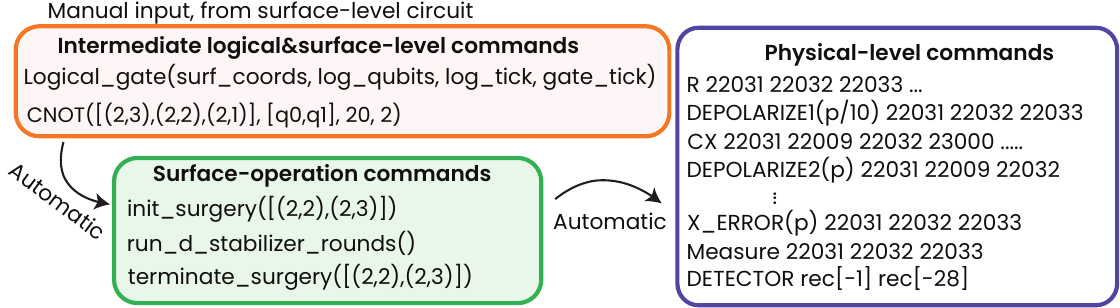}
 \caption{\label{fig:sim_SM} Simulation structure. An intermediate representation of the logical circuit with surface-level operations (manual input) is automatically transformed into surface-operation commands and then into a noisy physical-level schedule (e.g., depolarizing channels) with detector definitions (parity checks) for decoding.}
\end{figure}

To combine the two layers, we implemented a compiler that converts the surface-level circuit into a physical-level circuit. The compiler verifies correct gate implementation at the physical level. For example, each surface-level timestamp is realized as \(d\) stabilizer rounds. In addition, logical initializations (data-qubit initializations) occur at the beginning of a surface-level timestamp, and logical measurements (data-qubit measurements) occur at its end. Thus, the compiler assembles the physical-level circuit by invoking the appropriate routines from the physical-level code to implement the desired surface-level circuit.

\section{Reaching the factorization quantum circuit}
\label{app:circuit}
Shor’s factorization algorithm~\cite{shor1999polynomial} includes several steps, in which one of them can reach a super-polynomial speedup if implemented on a quantum computer. The algorithm for factoring a number \(N\) includes:

\begin{enumerate}
  \item Choose a random number \(1<a<N\) such that \(\gcd(a,N)=1\) (i.e., \(a\) and \(N\) are coprime).
  \item Use a quantum computer to find the (unknown) period \(r\) of the function
  \[
    f(x)=a^x \bmod N \quad\text{so that}\quad f(x)=f(x+r)=a^{x+r}\bmod N.
  \]
  \item If \(r\) is odd, return to Step~1 and choose another \(a\). If \(r\) is even, note that
  \((a^{r/2}-1)(a^{r/2}+1)=a^r-1\equiv 0 \pmod N\) and proceed.
  \item Compute \(d=\gcd\!\big(a^{r/2}-1,\,N\big)\). If \(d=1\) return to Step~1; otherwise \(d\) is a nontrivial factor (and so is \(N/d\)).
\end{enumerate}

The way a quantum computer is used to find the period proceeds in two steps. First, a circuit prepares the entangled state
\[
  \ket{\psi}=\frac{1}{2^{\,q-1}}\sum_{x=0}^{2^q-1}\ket{f(x),\,x}\,,
\]
which entangles the \(q\) ``input'' qubits encoding \(x\) (in binary) with ancillary qubits encoding \(a^x \bmod N\) (in binary). Second, the circuit applies the inverse quantum Fourier transform (QFT). Due to phase kickback, measuring the input register yields outcomes whose binary value divided by \(2^q\) equals \(s/r\) with nonzero probability, for \(s=0,1,\dots,r-1\) (each with equal probability).

The first part of the circuit can be executed schematically with \(q\) qubits satisfying \(N^2\le 2^q<2N^2\), and \(\lceil \log_2 N\rceil\) ancillary qubits to span all numbers up to \(N\), with \(q\) controlled unitaries. For factoring \(N=21\) with \(a=4\), the state to prepare is
\[
  \ket{\psi}=\ket{4^x \bmod 21,\,x}
  =\ket{1,0}+\ket{4,1}+\ket{16,2}+\ket{1,3}+\ket{4,4}+\ket{16,5}+\ket{1,6}+\ket{4,7}+\cdots
\]
showing that \(4^x \bmod 21\) has period \(r=3\). Thus it suffices to choose \(q=3\) (so \(x\in\{0,\dots,7\}\)), and the pre-QFT state can be written as
\[
\begin{aligned}
  \ket{\psi}={}&\ket{00001,000}+\ket{00100,001}+\ket{10000,010}+\ket{00001,011}\\
              &+\ket{00100,100}+\ket{10000,101}+\ket{00001,110}+\ket{00100,111}\,,
\end{aligned}
\]
where the comma inside the ket state description is  for readability. Since only three ancillary values appear, two ancillary qubits suffice. Mapping
\(\ket{00001}\mapsto\ket{00}\), \(\ket{10000}\mapsto\ket{11}\), and \(\ket{00100}\mapsto\ket{10}\), the state just before the inverse QFT becomes
\[
\begin{aligned}
  \ket{\psi}={}&\ket{00,000}+\ket{10,001}+\ket{11,010}+\ket{00,011}\\
              &+\ket{10,100}+\ket{11,101}+\ket{00,110}+\ket{10,111}\,.
\end{aligned}
\]
The circuit in Figure~1a prepares this state and performs the QFT.

Since \(r=3\), we would like probability \(1/3\) for outcomes corresponding to \(0,\,1/3,\) and \(2/3\). With three qubits, \(0\) is \(\ket{000}\), while the closest binary fractions to \(1/3\) and \(2/3\) are \(\ket{011}\,(3/8)\) and \(\ket{101}\,(5/8)\), respectively; other outcomes have nonzero probability. The histogram probabilities are
\[
  (0.344,\,0.015,\,0.063,\,0.235,\,0.031,\,0.235,\,0.063,\,0.015)
\]
for logical outcomes \(000,001,010,011,100,101,110,111\). This distinctive distribution was useful for validating correctness of our logical circuits when converting it from the original one from  Fig.~\ref{fig:illustration}a to the surface-compatible shown in Fig.~\ref{fig:surface_level}a.

To reach the maximal allowed error in our circuit, we note that the goal of Shor's algorithm is to measure the most-probable values in the circuit's final probability distribution. Within our circuit's distribution, the outcomes \(000,011,101\) have the largest probabilities,
\(0.344,0.235,0.235\), respectively, whereas all remaining outcomes occur with
probability at most \(0.063\). We refer to \(\{000,011,101\}\) as the heavy
outcomes and to the remaining bitstrings as light outcomes. To quantify the
deviation of the implemented circuit from this ideal output distribution
\(p = \{p_z\}\), we define the circuit error as the total-variation distance
between \(p\) and the realized distribution \(q = \{q_z\}\),
\begin{equation}
  \varepsilon_{\mathrm{circ}}
  \;:=\;
  \frac{1}{2}\sum_{z \in \{0,1\}^3} \bigl|p_z - q_z\bigr|.
\end{equation}
The minimal ideal separation between any heavy outcome \(h\) and any light
outcome \(\ell\) is
\(\Delta_{\min} = \min_{h \in H,\;\ell \in L} (p_h - p_\ell)
= 0.235 - 0.063 = 0.172\).
For any distribution \(q\) with total-variation error
\(\varepsilon_{\mathrm{circ}}\), the difference between a heavy and a light
probability can be reduced by at most \(2\varepsilon_{\mathrm{circ}}\), so
\begin{equation}
  q_h - q_\ell \;\ge\; \Delta_{\min} - 2\varepsilon_{\mathrm{circ}}
  \qquad \forall\, h \in H,\;\ell \in L.
\end{equation}
Consequently, any implementation satisfying
\begin{equation}
  \varepsilon_{\mathrm{circ}} < \frac{\Delta_{\min}}{2} \approx 0.086
\end{equation}
is guaranteed, in the worst case over all noise consistent with this
total-variation error, to preserve the strict ordering
\(q_{\text{heavy}} > q_{\text{light}}\) for all heavy–light pairs. In this
sense, a circuit error below about \(8.6\%\) is sufficient to keep the three
target outcomes statistically dominant over all others.

\end{document}